\documentclass{ceab}   
\usepackage{amsmath}
\usepackage{upgreek}
\usepackage{epsfig}     
\usepackage{graphicx}   
\usepackage{url}
\usepackage{xcolor}
\usepackage{natbib}     
\usepackage[T1]{fontenc} 
\usepackage{babel}       
    
\def\runningtitle{Owen et al.: Multimessenger backgrounds from galaxies}   
\def\articlenumber{7}
\def\ppages{1--4}

\begin{document}

\title{High energy extragalactic multimessenger backgrounds from starburst and dead galaxies}

\author{Ellis R. Owen$^{1,2}$\thanks{Speaker}, Yoshiyuki Inoue$^{1,3,4}$, 
Tatsuki Fujiwara$^{1}$ and Albert K. H. Kong$^{5}$ 
\vspace{2mm}\\
\it $^{1}$Theoretical Astrophysics, Department of Earth and Space Science, \\ 
\it Graduate School of Science, The University of Osaka,\\ 
\it Toyonaka 560-0043, Osaka, Japan  \\ 
\it $^{2}$Astrophysical Big Bang Laboratory (ABBL), RIKEN Pioneering Research Institute, \\
\it Wak\={o}, Saitama, 351-0198 Japan \\
\it $^{3}$Kavli Institute for the Physics and Mathematics of the Universe (WPI), \\
\it UTIAS, The University of Tokyo, Kashiwa, Chiba 277-8583, Japan \\
\it $^{4}$Interdisciplinary Theoretical \& Mathematical Science Program (iTHEMS), \\ 
\it RIKEN, 2-1 Hirosawa, Saitama 351-0198, Japan \\
\it $^{5}$Institute of Astronomy, National Tsing Hua University, \\
\it No. 101, Section 2, Kuang-Fu Rd, Hsinchu 30013, Taiwan (ROC)
}

\maketitle

\vspace{-1cm}
\begin{abstract}
Starburst galaxies are $\gamma$-ray sources. Canonically, 
their emission is driven by hadronic cosmic rays (CRs) interacting with interstellar gas, forming $\gamma$-rays via the decay of neutral pions. 
Charged pions are also formed in this process.  
They decay into secondary leptons, including electrons and neutrinos.  Starburst galaxies are therefore also expected to be neutrino sources, and  
their high-energy $\gamma$-ray emission may include a secondary leptonic component. 
Leptonic $\gamma$-rays may also 
originate from electrons directly energized by shocks within the interstellar medium of galaxies, or from pulsars and their surrounding halos. In the Milky Way, pulsars/pulsar halos are the dominant $\gamma$-ray source class. They are associated with stellar remnants or old stellar populations, and are presumably abundant in old galaxies. In this work, we 
show that the collective high-energy emission from galaxies can account for only a 
fraction of extragalactic neutrinos, but can form 
a major component of the extragalactic $\gamma$-ray background. Contrary to the traditional view, a substantial fraction of this radiation may originate from leptonic processes, including from old, quiescent galaxies. 
\end{abstract} 

\keywords{Cosmic rays - $\gamma$-rays - neutrinos - galaxies - millisecond pulsars/halos}

\sloppy

\vspace{-0.2cm}
\section{Introduction}
\label{sec:introduction}

\vspace{-0.2cm}
Recent advances in the past decade have greatly improved our understanding of the origins of 
high-energy photons and particles from beyond our galaxy. Key developments include new 
measurements of the diffuse neutrino flux by IceCube, spanning 30 TeV to PeV energies \citep{IceCube2013Sci, Aartsen2016ApJ, IceCube2020PhRvL}, and the extragalactic $\gamma$-ray background observed by \textit{Fermi}-LAT \citep{Ackermann2015ApJ}. 
These backgrounds can be decomposed into two components: one arising from emission by resolved, individual extragalactic sources, and another from the combined emission of all unresolved sources across the observable Universe. The unresolved component is believed to predominantly originate from populations of distant, point-like sources that cannot be individually detected \citep{Ackermann2015ApJ}. These include 
objects such as blazars \citep[e.g.][]{Inoue2009ApJ}, radio galaxies \citep[e.g.][]{Inoue2011ApJ...733...66I,Stecker2019ApJ} and 
starburst galaxies \citep[e.g.][]{Roth2021Natur, Chen2025arXiv}.

Starburst galaxies are cosmic ray (CR) reservoirs, with CRs mainly supplied by the end-products of stellar evolution, such as supernova remnants. Their interstellar conditions allow them to operate effectively as calorimeters, efficiently converting their CR supply into high energy $\gamma$-rays and neutrinos. This makes them well-motivated candidates for the diffuse extragalactic background flux, with nearby starburst galaxies found to be $\gamma$-ray bright \citep{Ajello2020ApJ}.  The contribution from starburst galaxies has been estimated to account for a significant fraction of the extragalactic $\gamma$-ray background 
and, in some model configurations, may even saturate it \citep{Roth2021Natur}. While their role in producing the extragalactic neutrino flux is likely to be less substantial, 
detailed spectral modeling variations may allow for higher contributions. 
Observations have also revealed $\gamma$-ray emission from nearby quiescent galaxies, such as M31 \citep{Ajello2020ApJ}. In these systems, the emission may be linked to evolved stellar populations and their compact high-energy sources, including millisecond pulsars (MSPs)
and TeV $\gamma$-ray-emitting pulsar halos.  
These sources could also contribute to the high-energy $\gamma$-ray background \citep{Xu2022PhRvD}. 

In this work, we present a new determination of the multimessenger high-energy emission from galaxies to systematically investigate the combined origins of the extragalactic $\gamma$-ray and neutrino backgrounds from galaxies. Compared to previous studies, our approach extends beyond starbursts to include main-sequence and quiescent galaxies. This allows for a more robust assessment of the high-energy emission originating from diverse galaxy types over a broad range of redshifts. 

\vspace{-0.2cm}
\section{Cosmic rays in galaxies}
\label{sec:CRs_in_galaxies}

\vspace{-0.2cm}
\subsection{Propagation and confinement}
\label{sec:prop_conf}

\vspace{-0.2cm}
CRs are typically energized within galaxies by shocks and turbulence. These are often driven by star-formation activities through feedback from supernovae (SNe) and winds. We therefore parameterize the CR supply rate within a galaxy using its SN event rate, and adopt a power-law injection spectrum with an index of -2.1, and an exponential cut-off of 50 PeV. 2 per-cent of the total CR energy is supplied to relativistic electrons, with the remaining fraction supplied to protons. Once energized, CRs diffuse through turbulent interstellar magnetic fields. They then slowly leak out of a galaxy or are attenuated by particle interactions. To account for CR propagation and losses, we adopt the model introduced by \citet{Owen2019A&A}. This invokes an interstellar medium (ISM) permeated by Kraichnan-type magnetic turbulence, where the magnetic energy density is set to be the same as the energy supplied by SNe over an advection timescale \citep[this is similar to the approach adopted by][]{Chen2025arXiv}.\footnote{An adjustment is made to ensure our model reproduces the magnetic field strength of the Galactic ISM for parameter values appropriate for the Milky Way.} 

\vspace{-0.2cm}
\subsection{Interactions and multimessenger emission signatures}
\label{sec:interactions_mm}

\vspace{-0.2cm}
Above a threshold energy of $\sim$280 MeV, hadronic CRs interact with ambient gas to produce charged and neutral pions. These pions decay, producing $\gamma$-rays, neutrinos and secondary electrons and positrons (hereafter, we do not distinguish between electrons and positrons). We model the production of these secondary particles using 
{\tt Aafrag} 2.01 \citep{2023CoPhCKachelriess}. 
This approach allows the direct computation of the pion decay $\gamma$-ray and neutrino spectra for a given galaxy. Hadronic interactions attenuate the primary CR flux within galaxies. This attenuation is accounted for self-consistently in our adopted CR distribution model \citep{Owen2019A&A}. 
Additional $\gamma$-ray emission arises from bremsstrahlung and inverse Compton scattering of CR electrons within a galaxy. In our calculations, we include contributions from both primary CR electrons and secondary electrons produced in hadronic collisions. To compute this emission, we consider that the electron distribution is held at a steady state where  
particle injection is balanced by cooling through radiative and collisional processes. In our model, both hadronic and leptonic losses, as well as their associated emission, depend on several key factors: magnetic field strength, gas density, and the intensity of the interstellar radiation field. These quantities are parameterized using the star formation rate. 

\vspace{-0.2cm}
\section{Pulsar populations in galaxies}
\label{sec:pulsar_populations}

\vspace{-0.2cm}
Pulsars and their surrounding halos are the dominant $\gamma$-ray source class in our Galaxy \cite[for a recent review, see][]{2022NatAsLopezCoto}, 
and are considered capable of driving most of the $>$500 GeV emission from other similar main-sequence galaxies, including M31 \citep{Xu2022PhRvD}.  
They are 
separated into two distinct populations based on their spin-down properties. The majority are canonical pulsars, which can host spatially-extended `pulsar halos' observed in TeV $\gamma$-rays.   
Pulsar halos are predominantly leptonic, diffusion-dominated systems that typically feature an extended region where CR diffusion is suppressed compared to the ISM average \citep[e.g.][]{Schroer2023PhRvD}. 
We construct a simplified pulsar halo spectral model using a single suppressed, uniform diffusion coefficient (reduced by a factor of 10 relative to typical ISM values) and a power-law CR injection spectrum with an exponential cutoff at 1 TeV and index of -1.6, gauged empirically from the spectrum of the Geminga pulsar halo when adopting a steady-state balance between electron injection and radiative cooling. 
We adopt this spectral model as a baseline for the combined emission from a population of pulsar halos in a galaxy, parameterized by their time-evolving spin-down rates.
  
We construct a population model to describe the evolution of pulsar halos in a galaxy. This considers that the population of pulsar halos at a given time in a galaxy can be characterized in terms of their birth rate, death rate, and spin-down rate, where the emission properties of each pulsar halo are parameterized by its spin-down luminosity. 
The number of pulsar halos in a galaxy with luminosities between  $L_{\gamma}$ and $L_{\gamma} + {\rm d}L_{\gamma}$ is $n_{\rm ph}(L_{\gamma}, t) {\rm d}L_{\gamma}$. The evolution of the halo population's luminosity function is then: 
\begin{align}
\frac{\partial n_{\rm ph}(L_{\gamma}, t)}{\partial t} + \frac{{\rm d}\;\!L_{\gamma}}{{\rm d}\;\!t} & \frac{\partial n_{\rm ph}(L_{\gamma}, t)}{\partial L_{\gamma}} = -\kappa \;\! n_{\rm ph}(L_{\gamma}, t) + f_{\rm ph}(L_{\gamma},t) \ ,
\label{eq:evo_eq}
\end{align}
where $\kappa$ 
represents  
their `death' rate, $f_{\rm ph}(L_{\gamma}, t)$ is the formation rate, and   
${{\rm d}L_{\gamma}}/{{\rm d}\;\!t}$ represents their fading, brightening, or possible re-brightening. 

The pulsar `birth' term is regulated by the SN event rate of a galaxy, with a fraction of SN events producing pulsar and halo systems. 
For the intrinsic birth luminosity function, we adopt a power-law distribution, which has been shown to accurately represent the initial spin-down (and hence luminosity) distribution of the $\gamma$-ray and radio pulsar populations in the Milky Way \citep[see][]{Watters2011ApJ}. The fading of pulsar halos over time is modeled by tracking the evolution of the spin-down luminosity of their parent pulsars. 
The halo death rate is modeled using a 
characteristic death timescale. This is derived from the empirical pulsar `death line’, considered to represent the threshold for sustaining electron-positron pair production within the pulsar's magnetosphere. 

The second population are MSPs. These are 
recycled systems originating from evolved compact binaries that have been spun-up through past accretion episodes or dynamical interactions  \citep{Alpar1982Natur, Hui2010ApJ}. 
The $\gamma$-ray spectrum of MSPs is similar to that of normal pulsars. As a characteristic MSP spectral model, we adopt a log-parabola with  parameter values fit by \cite{Lloyd2024MNRAS} for the stacked emission from 118 \textit{Fermi}-detected MSPs. We do not include a halo component in the spectrum. 
The MSP population is modeled in a similar way to the pulsar halos, characterized in terms of their birth and death rates in a galaxy, and their ageing (fading).  

While the death condition for MSPs is the same as that for canonical pulsars, their far lower spin-down rate results in a relatively constant $\gamma$-ray luminosity over Gyr timescales so that the fading term in our MSP population model is negligible. MSP formation depends on a spin-up accretion process or dynamical interactions, indicating that these systems form as binaries \citep{Alpar1982Natur} or in regions of high stellar densities \citep[e.g.,][]{Hui2010ApJ}. The birth rate of MSPs $f_{\rm msp}(L_{\gamma}, t)$ is therefore set by the number of SN events from a fraction of progenitor stars which have sufficient mass to yield a neutron star, and is then scaled by  
the relative frequency of dynamical encounters.  
The evolution of a population of MSPs in our model is then described by: 
\begin{align}
\frac{{\rm d} n_{\rm msp}(L_{\gamma}, t)}{{\rm d} t} & = - \kappa \;\! n_{\rm msp}(L_{\gamma}, t) +  f_{\rm msp}(L_{\gamma}, t) \ . 
\label{eq:birthdeath_MSP}
\end{align} 
\citep[see also][]{Wu2001PASA}. 
By directly integrating this equation and assuming (1) there are initially no MSPs, and (2) the MSP death timescale is much longer than the evolutionary timescale of a galaxy, the result reduces to a time-independent luminosity function, $n_{\rm msp}(L_{\gamma}) \approx f_{\rm msp}^0(L_{\gamma}) \;\! M_{\rm *}$, where $M_{\rm *}$ is the stellar mass of a galaxy and $f_{\rm msp}^0(L_{\gamma})$ is the initial MSP luminosity function normalized to the formation rate of MSPs per stellar mass.\footnote{This is estimated from a binary formation rate of 0.7 \cite[see][for a review of multiplicities of massive stars]{Duchene2013ARA&A} 
and enhanced by the 
average stellar interaction rate (given by the 
square of the stellar density; see e.g.~\citealt{Hui2010ApJ}).} We model $n_{\rm msp}(L_{\gamma})$ using a broken power-law with best-fit parameters from 
\cite{Bartels2018MNRAS}. 

\vspace{-0.2cm}
\section{Multimessenger emission from galaxies}
\label{sec:multimessenger_galaxies}

\vspace{-0.2cm}
\subsection{Prototype galaxy model}
\label{sec:prototype}

\begin{figure}[h]
\centering
\vspace*{-0.2cm}
\includegraphics[scale=0.5]{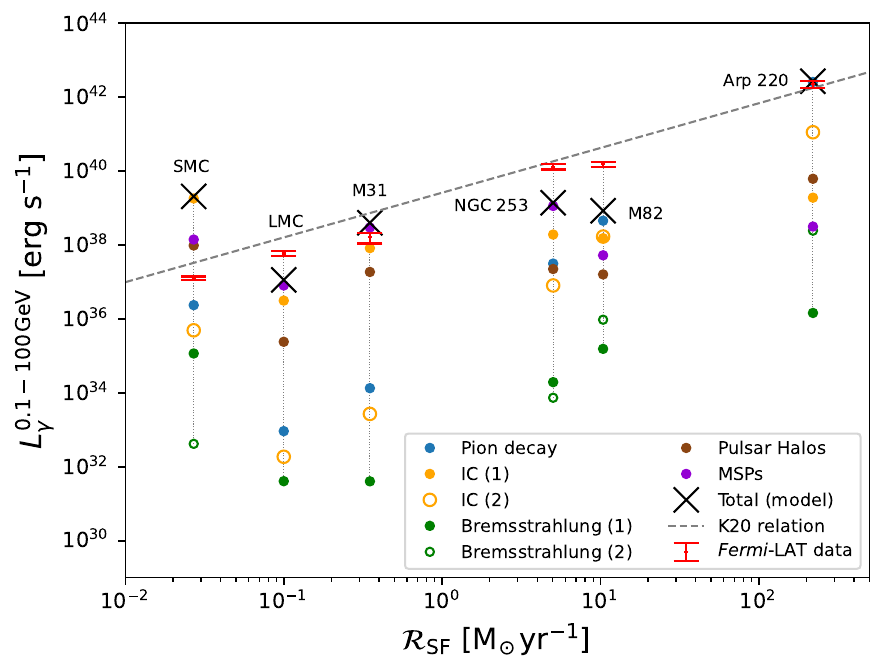}
\vspace*{-0.2cm}
\caption{Relationship between 0.1-100 GeV $\gamma$-ray luminosity and star-formation rate for six nearby galaxies. Total $\gamma$-ray luminosities are shown by red (observed) and black (model predicted) crosses. 
Contributions from different emission components are indicated by circles, 
with emission associated with primary (1) / secondary (2) CRs indicated by filled / open circles. The empirical relation of \cite{Kornecki2020A&A} is shown in gray (K20).} 
\label{fig:gamma_lum}
\vspace*{-0.2cm}
\end{figure}

\noindent
By parameterizing the emission models described in Sections 2 and 3, we construct a prototype model for the multimessenger high-energy emission from a galaxy. The model is specified by 
four inputs: a galaxy's effective radius, stellar mass, star-formation rate, and redshift. Other parameters, such as CR spectral properties, are either fixed or derived from these inputs. 
We validate our model against the spectra of six nearby galaxies where the $\gamma$-ray emission is believed to be dominated by CR activity associated with their stellar populations \citep{Ajello2020ApJ}. Without detailed fitting, our model predictions show reasonable agreement with these observations. Some differences emerge in the predicted total 0.1-100 GeV $\gamma$-ray luminosities when compared to \textit{Fermi}-LAT data, as shown in Figure \ref{fig:gamma_lum}. These are primarily driven by variations in the galactic magnetic field, which may include components not accounted for in our model. Both CR confinement and inverse Compton emission (which is particularly sensitive to the efficiency of electron synchrotron cooling) are influenced by the magnetic field. Despite these differences, our model reproduces the empirical trend (K20), indicating that it is able to capture the overall behavior of $\gamma$-ray emission across a population of galaxies.

\vspace{-0.2cm}
\subsection{Galaxy population model}
\label{sec:universemachine}

\vspace{-0.2cm}
We adopt the empirical semi-analytic galaxy-halo connection simulation, {\tt UniverseMachine} \citep{Behroozi2019MNRAS} as a galaxy population model. This 
provides a self-consistent view of galaxies' star formation rates, stellar masses, host halo properties, assembly histories, and redshifts. 
It is constrained by 
empirical quantities, including galaxies' observed stellar mass functions, star-formation rates, quenched fractions, ultra-violet (UV) luminosity functions, UV-stellar mass relations and 
the environmental dependence of quenching. 
These empirical constraints ensure that {\tt UniverseMachine} offers a robust baseline model, calibrated against the real Universe. 

\vspace{-0.2cm}
\subsection{Multimessenger backgrounds}
\label{sec:results}

\begin{figure}[h]
\centering
\includegraphics[scale=0.34]{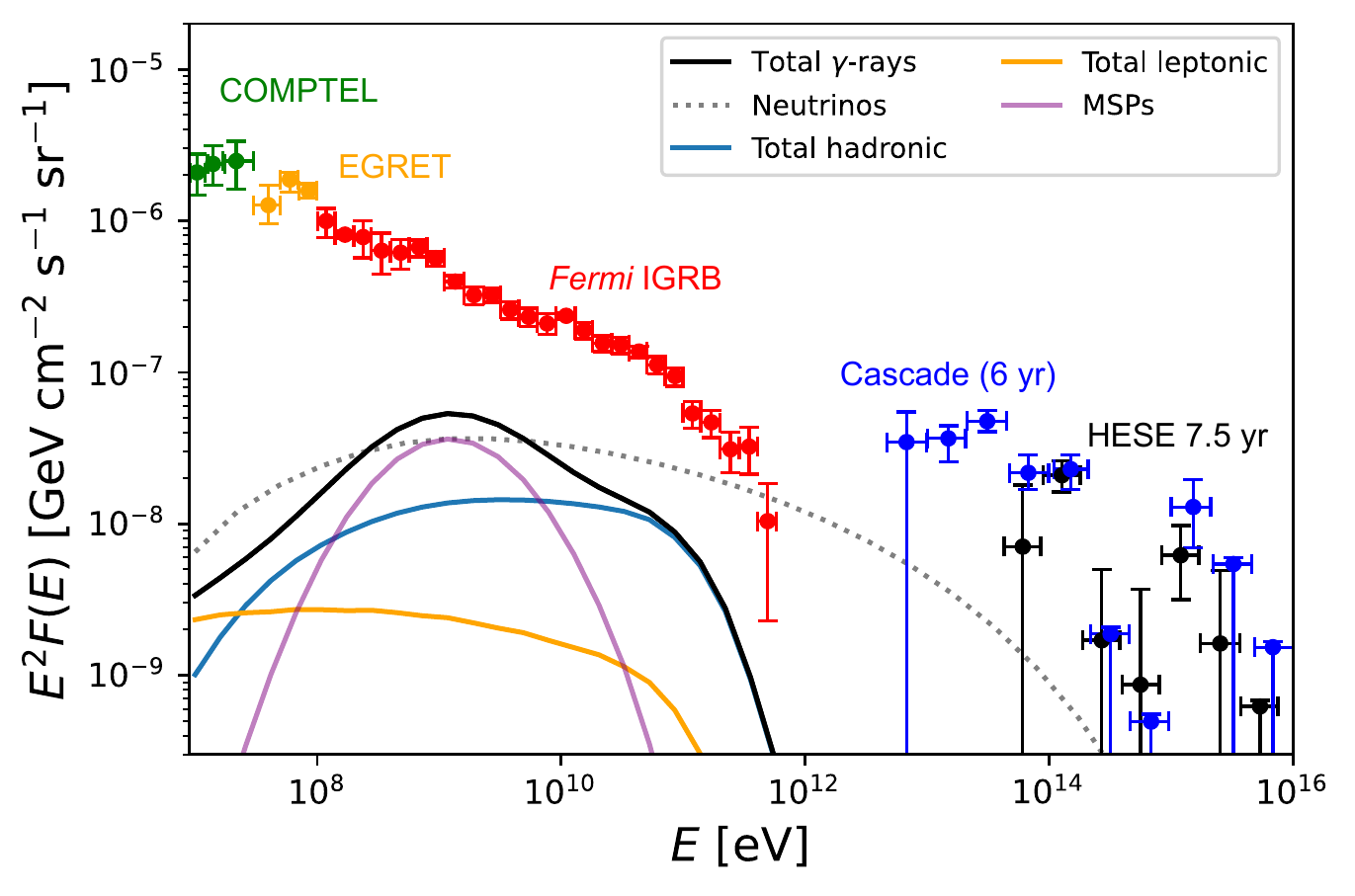}
\vspace*{-0.2cm}
\caption{Galaxy contributions to the diffuse extragalactic $\gamma$-ray and all-flavor neutrino flux, compared with observational data from \textit{COMPTEL} \citep{Kappadath1996A&AS}, \textit{EGRET} \citep{Sreekumar1998ApJ}, \textit{Fermi}-LAT \citep{Ackermann2015ApJ} and IceCube \citep{IceCube2013Sci, Aartsen2016ApJ, IceCube2020PhRvL}. The total predicted contributions are shown in black, as labeled. The $\gamma$-ray flux is comprised of a mix of MSP emission, pion decays, and a sub-dominant leptonic inverse Compton component (attributed mainly to secondary CR electrons). 
}
\label{fig:EGB_final}
\vspace*{-0.3cm}
\end{figure}

\vspace{-0.2cm}
We post-process the {\tt UniverseMachine} outputs with our galaxy prototype model to calculate the combined $\gamma$-ray and all-flavor neutrino emission from galaxy populations. Using a cosmological radiative/particle transfer formulation, we determine the flux arriving at Earth, ensuring conservation of the phase-space density of photons and neutrinos in an evolving cosmology. We integrate over all galaxies within the redshift range from 
z=0 to 6, beyond which their flux contribution becomes negligible. 
Our calculations account for the attenuation and cascade reprocessing of high-energy $\gamma$-rays through pair-production interactions with intergalactic radiation fields. These include the cosmic microwave background (CMB) and the extragalactic background light (EBL; adopting the model of \citealt{Saldana-Lopez2021MNRAS}). Figure \ref{fig:EGB_final} presents our results for the total $\gamma$-ray and all-flavor neutrino emission contributions from galaxies.

\section{Remarks and conclusions}
\label{sec:remarks_conclusions}

\vspace{-0.2cm}
We find that 
most of the neutrinos in our model originate from CR interactions in strongly starbursting galaxies. 
However, our calculations indicate that 
it is unlikely these CR processes in galaxies alone could saturate the neutrino flux under any plausible model configuration. Instead, much of the remaining observed neutrino flux likely originates in other environments, e.g. blazars and Seyfert galaxies.   

Our calculations also show that the $\gamma$-ray emission from CR processes in galaxies 
likely represents a non-negligible component of the extragalactic $\gamma$-ray background. 
In contrast to some previous works \citep[e.g.][]{Roth2021Natur}, our results suggest that galaxies do not saturate the $\gamma$-ray background at any energy. However, they may still contribute a substantial component of the flux at energies above a few tens of GeV up to 1 TeV. We also confirm the presence of a leptonic $\gamma$-ray emission component from galaxies at energies below $\sim$0.1 GeV reported by previous studies \citep[e.g.][]{Roth2021Natur, Chen2025arXiv}, and additionally find that quiescent `dead' galaxies with old stellar populations could contribute substantial GeV $\gamma$-ray emission Gyrs after their main star-formation activity. This introduces evolved galaxies as a potential source population in the GeV $\gamma$-ray sky that has not previously been widely considered. Their contribution at these energies could be comparable to, or even exceed, that of starburst galaxies and is strongly dominated by MSP populations. This may naturally explain a spectral excess seen in \textit{Fermi} data at around 1 GeV.

\vspace{0.1cm}
\footnotesize
\noindent
\textbf{Acknowledgements:} E.R.O is an international research fellow under the Postdoctoral Fellowship of the Japan Society for the Promotion of Science (JSPS), supported by JSPS KAKENHI Grant Number JP22F22327, and also acknowledges support from the RIKEN Special Postdoctoral Researcher Program for junior scientists. 
Y.I. is supported by an NAOJ ALMA Scientific Research Grant 2021-17A, JSPS KAKENHI Grants JP18H05458, JP19K14772, and JP22K18277, and the World Premier International Research Center Initiative (WPI), MEXT, Japan. This work was partially supported by the joint research program of the Institute for Cosmic Ray Research, the University of Tokyo. 
Numerical computations were carried out on the Cray XC50 supercomputer at the Center for Computational Astrophysics, National Astronomical Observatory of Japan and at the Yukawa Institute Computer Facility, Kyoto University. We thank Dr. Alison Mitchell for providing a helpful and constructive review that improved this article. 

\vspace{-0.4cm}

\bibliographystyle{apalike}
\bibliography{references}

\end{document}